# Chip-to-chip ODDM network with optically enabled equalization

Andrea Zazzi and Jeremy Witzens, *Senior Member, IEEE*

*Abstract*—We propose and model an optical communication scheme for short distance datacom links based on the distribution of information across a wide comb spectrum. This modulation format, orthogonal delay division multiplexing, allows the multiplexing of data streams from multiple modulators, as well as the deserialization and equalization of the data in the optical domain. A concrete communication system, that allows the transport of 400 Gb/s across a single CWDM channel with a single 80 GHz cutoff lithium niobate on insulator modulator, is modeled under consideration of all noise sources present in the system and its sensitivity to group velocity dispersion is analyzed. Data is deserialized and equalized at the receiver with a 5-tap optical equalizer. This communication architecture may provide a path forward to implement high-baud-rate signaling in short-reach optical links without requiring high-speed ADCs and electronic deserializers at the receiver, thus maintaining the in-package power consumption at manageable levels.

*Index Terms*—Co-packaged optics, in-package optics, optical computing, optical fiber communication, optical interconnects, optical modulation, optical pulse shaping, optical receivers, optical signal processing, optical transmitters, silicon photonics

## I. INTRODUCTION

THE rapid growth in the processing capabilities of hardware accelerators for deep neural networks (DNNs), in particular graphics processing units (GPU), has led to bottlenecks at the chip inputs/outputs (I/O) and to a rekindled interest in optical I/Os at the chip and package level. I/O requirements for single chips now approach a terabit per second [1]. A number of strategies have been devised to implement in-package optics in a compact, cost- and power-effective way. While interconnects at centralized network hubs such as top-of-the-rack (ToR) switches have been moving to increasingly high data rates, some of the solutions proposed for chip-to-chip interconnects aim at massive parallelization of comparatively low-speed optical channels, by means of both fiber parallelization and the wavelength division multiplexing (WDM) of an increased number of channels [2]. The rationale there is to avoid the power overhead associated to the serialization/deserialization (SerDes) required to generate high-baud-rate signal streams from the much lower internal clock-speeds of the digital electronics. To make the resulting large number of electro-optic (E-O) modulators power efficient, and to be able to fit them on a compact chiplet, it is then necessary to build the system around ultra-compact and very low power modulators such as micro-ring modulators (MRMs). While these are by now very well understood devices [3] that have been implemented in a number of platforms, including zero-change complementary metal-oxide-semiconductor (CMOS) processes [4], they are also very sensitive to thermal fluctuations and should ideally be trimmed [5] to reduce the power overhead associated to their resonances' alignment with the carrier wavelengths.

On the other hand, there has been tremendous progress made in Pockels-effect based Mach-Zehnder modulators (MZM) in recent years. Thin film lithium niobate on insulator (LNOI) stands out, with devices featuring low optical losses, very high bandwidths in excess of 100 GHz, reduced drive voltages, and high optical power handling capability [6]. Progress has also been made in the heterogeneous integration of such modulators into the silicon photonics platform [7]. Some more recently explored material systems such as barium titanate (BTO) even offer the prospect of being integrated at the level of 200 mm wafers [8]. While such modulators are typically travelling wave (TW) devices and thus have a relatively high energy consumption associated to their characteristic impedance, this energy consumption can be competitive even at short distances if spread over a sufficiently large amount of data, i.e., if the high baud-rate achievable with these devices is being fully exploited. To give a numerical example, a TW-MZM with a 50Ω impedance and driven with 2 V$_{pp}$ dual-drive signal consumes 80 mW, but this corresponds to only 200 fJ per bit if it is able to support 400 Gb/s, such as would be reached with 200 GBd 4-level pulse amplitude modulation (PAM-4). While such capabilities are still out of reach of current commercial deployments, 100 GBd PAM-4 has already been shown a few years back [9] and current driver electronics reach cutoff frequencies in excess of 70 GHz, for example implemented in SiGe heterojunction bipolar transistor (HBT) technology [10]. There can thus also be a case for using such modulators at short distances, provided the power overhead associated to serializing and deserializing to and from such high baud rates is reduced, in particular inside the package, as the heat-sinking capability of in-package optics is one of the major challenges that needs to be tackled to improve their scalability.

This work was supported in part by the Federal Ministry of Education and Research of Germany (BMBF), Custer4Future NeuroSys, under Grant 03ZU1106BA and by the German Research Foundation, priority program "Electronic-Photonic Integrated Systems for Ultrafast Signal Processing," under Grant 403188360. Corresponding author: Jeremy Witzens.

Andrea Zazzi (e-mail: azazzi@iph.rwth-aachen.de) and Jeremy Witzens (e-mail: jwitzens@iph.rwth-aachen.de) are with the Institute of Integrated Photonics of RWTH Aachen University, 52062 Aachen, Germany.

This is all the more important, as the case for high baud rates on the receiver (Rx) side is much weaker for in-package optics. While modulators and their drivers are one of the main sources of power consumption in optical transceivers, photodetectors (PDs) and transimpedance amplifiers (TIAs) typically consume much less power. Instead, the clock-data-recovery (CDR) and the equalizer [11] are typically a major burden on the overall power consumption of a high baud-rate Rx [12]. The buffers and track-and-hold amplifiers (THA) required in the analog-to-digital converter (ADC) front-end of such high-baud-rate systems also result in significant power consumption [13]. Moreover, while it is challenging to design modulator drivers into CMOS processes due to their small drive voltage capabilities [14], state-of-the-art PD+TIA supporting over 50 GBd PAM-4 have been fabricated in CMOS [15], so that there is an even stronger incentive there to stay within the capabilities of current CMOS processes.

In this paper, we explore the possibility of implementing a robust optical processing scheme to both deserialize and equalize a high-baud-rate data stream arriving at the Rx. With it, the high data rates in reach of state-of-the-art TW-modulators may be fully exploited without incurring the resulting deserializer power penalties at the Rx. Concretely, we look into deserializing an E-O bandwidth limited 200 GBd PAM-4 signal into four equalized 50 GBd PAM-4 data streams in line with current 800GAUI-8 electrical specifications with an optical system of manageable complexity.

Recently, we have introduced the concept of *orthogonal delay division multiplexing* (ODDM) in the context of on-chip broadcasting and processing of signals in electrical-optical-electrical (OEO) artificial neural networks (ANNs) and showed that it has a high robustness against environmental temperature drifts when implemented with a balanced network on a single chip [16]. ODDM transmission relies on frequency comb sources, but instead of mapping channels to individual comb lines as done in WDM systems [17]-[19], information is mapped over the entire comb spectrum while maintaining orthogonality between the channels. Here, we evaluate the application of this multiplexing scheme to the interconnection of digital hardware accelerators via fiber optics, satisfying the constraints and goals outlined above.

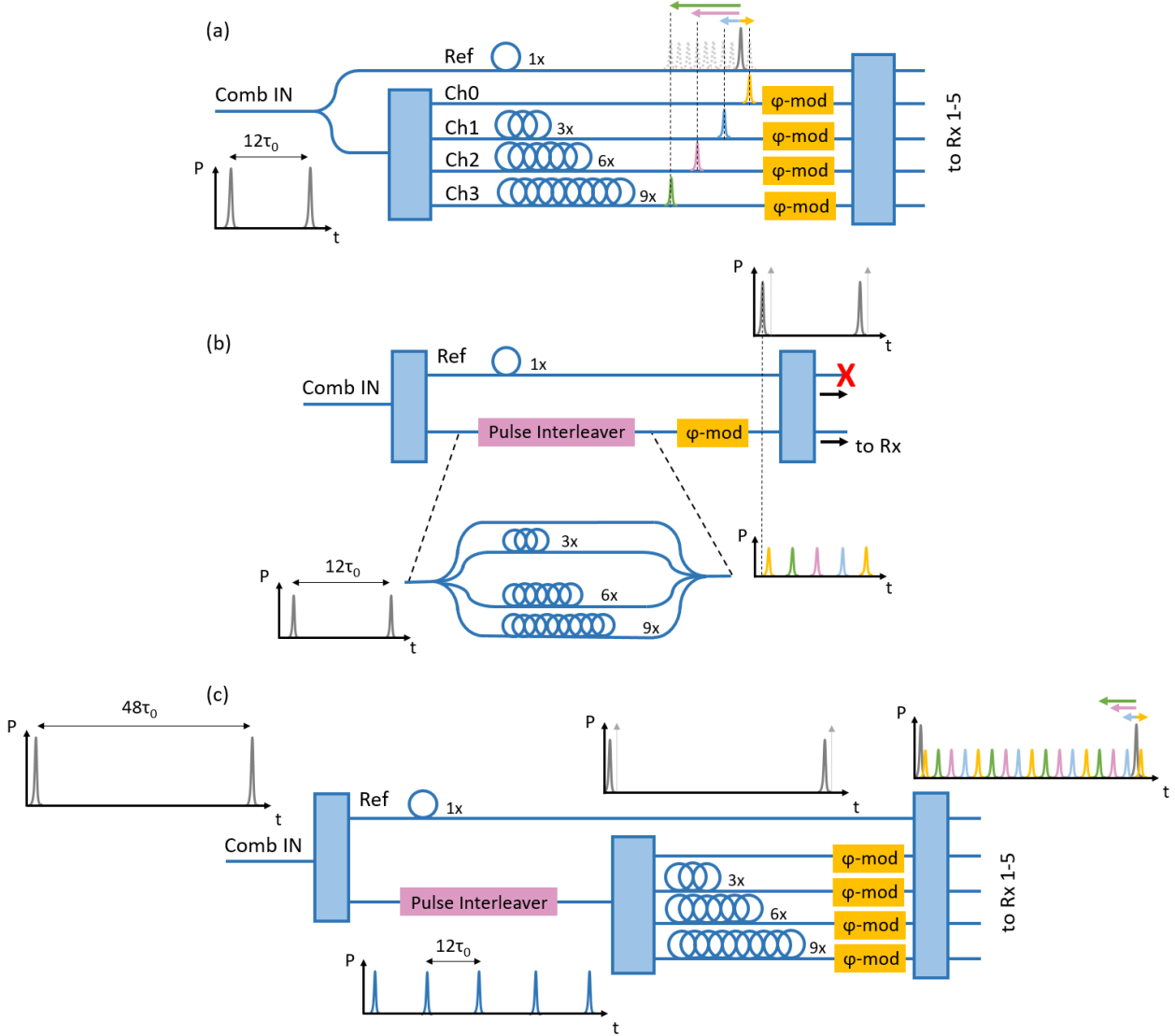

**Fig. 1.** ODDM Tx architectures. In (a), signals generated by 4 different phase-modulators are multiplexed on a single comb and broadcasted to 5 downstream Rxs. In (b), a single phase-modulator generates a signal with a baud rate reaching up to 4x the comb repetition rate. To avoid aliasing, a train of 4 interleaved comb pulses is first created. In (c), both aspects are combined. A wider spectrum with narrower pulses is required and the delays in the interleaver are increased to 0x, 12x, 24x and 36x $\tau_0$. Information is carried by 4 interleaved pulses in (a) and (b), that are color coded (the stronger grey pulse is the reference pulse for the self-homodyne detection scheme). In (c), the number of independent pulses is increased to 16. Here, color-coding indicates pulses that have been modulated by the same phase-modulator, but that correspond to different sampling times.

In Section II, we build up the system concept introducing networks that allow the multiplexing of data streams generated by multiple modulators, the deserialization and equalization of the data stream generated by a single modulator, or both. Section III describes how equalization is done in the optical domain and Section IV summarizes numerical modeling results. These focus on a single modulator link with built-in optical equalization. Appendices I-III put the architecture explored here in a broader theoretical framework and analyze its robustness against group velocity dispersion (GVD).

## II. SYSTEM CONCEPT

Figure 1(a) shows the diagram of a transmitter (Tx) system in which four data streams generated by separate phase-modulators are broadcasted to five separate receivers, each receiving the entire data stream from a unique single-mode fiber.

A comb source is first split between a reference branch carrying half the optical power and a second branch, that is additionally split between four independent modulators, as this allocation optimizes the optical power budget [16]. In addition, group delays are applied in each branch with a magnitude $\tau_0$ for the reference branch and $3n\tau_0$ for the modulated branches, wherein $n = 0 \ldots 3$ is an integer that indices the data stream (ODDM channel). After modulation, these five branches are sent through a combiner, each output of which carries the entire information and can be sent to an Rx via an optical fiber. In the Rx, demodulators tuned to a specific ODDM channel receive the corresponding data with delayed self-homodyne detection.

Earlier works have been reported on intensity modulated (IM) pulse interleaving in integrated photonics platforms used in conjunction with direct detection (DD) [20]. An important distinction here is that the additional reference branch allows downstream coherent detection, which increases robustness against GVD (see Appendices I and II) and allows optical sampling and signal processing at the Rx (see Section III).

As shown in Fig. 2, inside the demodulators, the light is split in two branches and delayed in such a manner that the pulse originating from the reference branch overlays in time with the modulated pulse that is to be demodulated, creating a differential photocurrent at the balanced photodetector pair (PBD) at the demodulator output. A low-speed phase shifter (PS) allows setting the demodulator at the quadrature point, such that the obtained differential photocurrent swing is maximized for a given amount of phase shift applied at the corresponding phase-modulator. While in the general case a differential photocurrent can result from either the reference pulse travelling through the upper demodulator branch and the modulated pulse travelling through the lower demodulator branch or the contrary, here the relative time delays are chosen such that only one such configuration leads to time overlay between pulses. This is visible in the timing diagrams of Fig. 2, in which the only pairs of coinciding pulses are the reference pulse in the lower branch (in grey) and a pulse from a modulated channel (in blue), exemplarily chosen to have the index $n = 1$, in the upper branch.

Since the combiner at the output of the Tx in Fig. 1(a) has five inputs, it can be implemented as a 5-by-5 multimode interferometer (MMI) without additional loss of power, so that the broadcasting over multiple Rxs can be implemented without further burdening the link budget. This is essentially the network topology already introduced in [16] for an on-chip ANN.

For the orthogonality between the modulated pulses to hold, some constraints have to be applied to the time delay $\tau_0$. If the pulses do not overlay in time at the point at which they are combined, orthogonality is obviously achieved. Since the pulse width is in the order of $1/\Delta\nu$, with $\Delta\nu$ the spectral width of the comb, the minimum $\tau_0$ has also to be in this order (the exact coefficient depends on the pulse shape and on the acceptable level of crosstalk). However, complete temporal separation, while sufficient, is not always necessary: Sinc-shaped pulses as resulting from a square shaped spectrum have an infinite extent. However, setting $\tau_0$ to $1/\Delta\nu$ then also results in orthogonality [21]. In the time domain, it can be inferred from the maximum of one pulse overlaying with zero-crossings of the others. As previously pointed out [16], this multiplexing scheme has some important differentiations relative to direct detection orthogonal time division multiplexing (OTDM). In particular, data can also be applied to fully dispersed and overlapping pulses, provided the subcarriers are delayed by the proper amounts in each of the branches. However, while this applies to the diagram shown in Fig. 1(a), applying data to undispersed pulses will be required to obtain the time-domain sampling functionality used in the following (Fig. 1(b)). Even then, an improved tolerance to GVD remains, since the system remains robust against pulse broadening and pulses overlapping after the sampling operation, provided the final pulse width remains less than the comb's repetition time (see Appendix I for the general theoretical framework and Appendix II for an in-depth mathematical analysis).

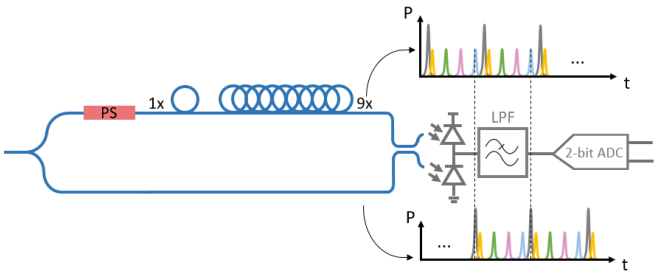

**Fig. 2.** Schematic of a fixed-channel demodulator (without equalization). After the first splitter, a delay is applied to the top branch in order to synchronize a modulated pulse with the reference pulse. After feeding these through a DCS, the resulting differential signal is transduced by a BPD, that is followed by an anti-aliasing filter or a switched integrator and a 2-bit ADC performing the thresholding operation for the PAM-4 reception. The PS in the upper arm biases the demodulator at quadrature for maximum signal strength.

In Fig. 1(b), interleaved pulses are not used to transport data from multiple modulators, but to sample a signal at a multiple of the comb repetition rate. In this case, the undispersed pulse train first needs to be converted into a train of interleaved pulses, in this example with four times the repetition rate. Since these pulses do not need to have equal phases, this does not require the extinguishing of comb lines and the associated optical power loss. In fact, such a pulse train can be straightforwardly obtained with the simple interferometer depicted in Fig. 3(a) [22]. The pulses need to stay on the general ODDM raster defined above, with time increments remaining multiples of $3\tau_0$, so that the reference pulse remains intercalated according to the scheme described above. This train of four pulses is then sent through a single modulator. At the Rx, the light is split four ways and each of the four pulse trains are received by a separate demodulator, with a differential delay adapted according to the ODDM channel number, and converted into a deserialized PAM-4 signal.

The interferometer shown in Fig. 3(a) is very robust, as each of its internal optical paths generates pulses with a different delay that consequently cannot interfere with each other. As a drawback, the interleaved pulse train is presented with equal power at both its outputs. This is not a disadvantage for an architecture as shown in Fig. 1(c) with multiple modulated branches, in which both outputs can be used. In the present case, Fig. 1(b), this would, however, lead to 3-dB losses. For this reason, the more general programmable pulse shaper (PPS) shown in Fig. 3(b) can be used. We have previously shown that such an interferometer allows the generation of an almost arbitrary superposition of pulses, in this case spaced by multiples of $3\tau_0$, with arbitrary pulse amplitudes and phases, provided the corresponding spectrum has comb line power levels below those of the comb lines at the input of the PPS [16]. Here, it allows us to generate an interleaved pulse train of constant amplitude (but irregular phase), while at the same time coupling it almost entirely to one interleaver output, as numerically verified by training it for that task. In more detail, it consists of four units, each unit comprising two PSs and two 3-dB directional-coupler-splitters (DCS) applying a 2-by-2 unitary matrix transformation to the field amplitudes transported in its two branches [23]-[25], as well as a delay loop introducing a group delay of $3\tau_0$.

The transformation that needs to be ideally applied by the PPS to the comb spectrum to obtain the interleaved pulse train [26] can be derived in analogy to the Talbot effect in MMIs. Following the formalism and the notations of [27], the target phase that needs to be applied to a specific comb line of index $q$ can be straightforwardly derived as

$$\frac{3L_\pi}{N}\frac{\nu(\nu+2)\pi}{3L_\pi} = \frac{\nu^2\pi}{N} + \frac{2\nu\pi}{N} \qquad (1)$$

where $N$ is the number of interleaved pulses and $\nu = \mathrm{mod}(q, N)$, with $q$ an index numbering the comb lines. The second term in Eq. (1) corresponds to a simple time delay and can be dropped. $\nu^2\pi/N$ is thus the phase that is ideally applied by the PPS.

While this structure is now truly an interferometer, in the sense that there are coherent interactions between the different paths followed by the light, it remains very robust against temperature fluctuations in terms of the power envelope of the generated pulse train: While each pulse is generated by light traveling over several optical paths, each of these contain an equal number of short and long (delay loop) path segments for a given pulse. As a consequence, the phase of the light changes with temperature after transmission through the PPS, but does so in the same way for every path contributing to a given pulse. The relative phase between different pulses drifts with temperature, but this can be straightforwardly compensated with a single control signal without having to retrain the PPS.

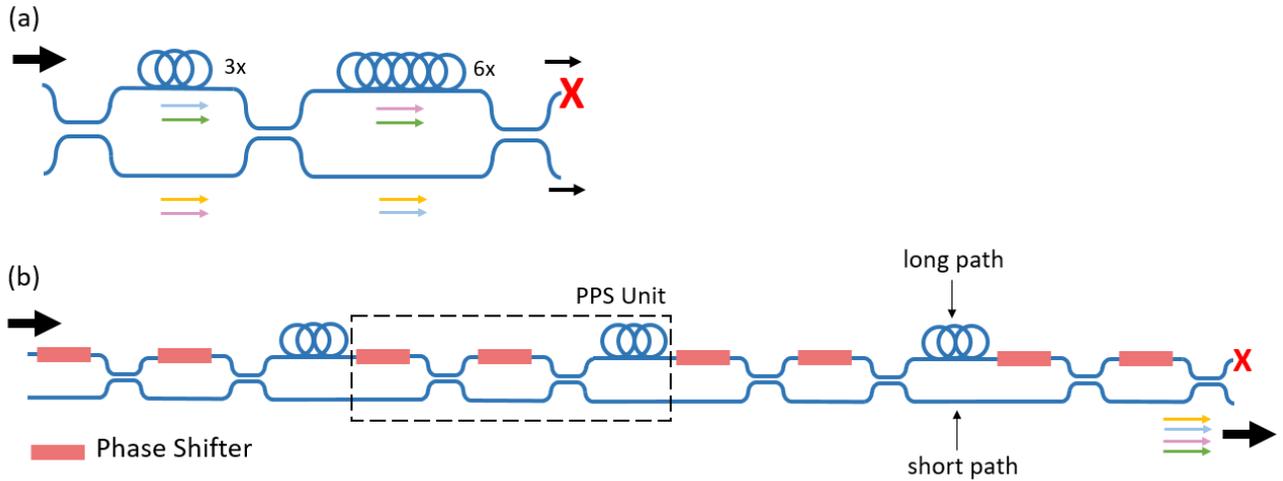

**Fig. 3.** (a) Simple pulse interleaver generating a train of 4 interleaved pulses at both its outputs and (b) more general programmable pulse shaper that allows the generation of a single interleaved pulse train at one of its outputs. This PPS is used for more general pulse shaping in the optical equalizers described in Section III and Fig. 4.

For this, it is sufficient to add after each delay loop a phase correction compensating for the temperature dependent phase difference between the long and shorter paths, which is identical for each of the PPS units. To avoid compensating for the bias point dependent nonlinearity of thermal phase shifters, this might be best implemented by separate PSs sharing a common bias point over the entire device.

Figure 1(c) finally combines the architectures of Fig. 1(a) and 1(b) into a system that multiplexes the data streams generated by 4 independent modulators, while at the same time allowing to oversample each of these by a factor 4, populating a total of 16 ODDM channels. For these to each fill different slots, the pulse interleaver is now adapted to space the pulses by $4 \times 3\tau_0 = 12\tau_0$, requiring a smaller $\tau_0$ at fixed comb repetition rate and a larger comb spectrum of at least 48 comb lines (more if the spectrum is not square shaped [16]).

Based on the high-level architectural aspects described so far, essential aspects relating to the optical power budget can already be discussed without a numerical model. The detection of a differential photocurrent with a BPD increases the transduced signal by 3 dB. This is, however, voided by the fact that modulated pulses and reference pulse are both split at the entrance of the demodulator, but each only contribute in one of its branches. This 3-dB loss is the price to pay for transporting both the signals and reference through the same fiber, an essential feature to maintain the relative phase stability across chips.

In the point-to-point link represented in Fig. 1(b), we incur a 3-dB penalty from only one of the two outputs of the DCS being used at the output of the Tx. A less evident penalty, compared to four parallel 100 Gb/s links provided with ¼ of the power (equal losses as the 1-to-4 splitter at the Rx sending the light to the four required demodulators), results from the fact that the modulated pulses are sent to all the demodulators, even though each demodulator recovers the data from only a single modulated pulse train, unless equalization is being applied. As a consequence, the power of the modulated pulses is reduced from what it could be if routed only to the relevant demodulator. The corresponding penalty scales with the square root of the number of utilized ODDM channels, due to the self-homodyne reception, since only the modulated pulses suffer from this and the reference pulse power is optimally used. This results in an additional $10log_{10}(\sqrt{4}) = 3$ dB for Fig. 1(b) and $10log_{10}(\sqrt{16}) = 6$ dB for Fig. 1(c). The cumulative effect is thus a 6-dB penalty for both Fig. 1(b) and Fig. 1(c) compared to conventional parallel single mode (PSM) or WDM links with a number of channels equating the number of ODDM channels utilized here. A few finer corrections arise from the fact that our architecture only suffers from half the modulator insertion losses, by nature of the coherent detection, but also suffers from increased shot noise resulting from the excess common-mode photocurrent generated at the BPD from off-target pulses that are not synchronized with the reference [16].

3 dB out of the 6-dB penalty derived above are a cost that is associated to an important benefit, namely the possibility of optically implementing an equalizer. For this, the 4 pulse trains transiting through a given modulator have to be available at a single demodulator, so that the associated penalty, $10log_{10}(\sqrt{4}) = 3$ dB, is necessary. Doing this with an IM/DD OTDM architecture would already have resulted in a 6-dB penalty by itself, since each pulse would, there too, have to be shared over 4 Rxs.

The other 3-dB excess link penalty arising in the architecture shown by Fig. 1(b) could be recovered by using two fibers to route both Tx outputs to the Rx and hooking up only two demodulators to each. This would reduce the data rate per fiber to 200 Gb/s, but maintain a 400 Gb/s operation at the modulator, so that the in-package Tx power consumption would remain unchanged.

In Fig. 1(c), on the other hand, the other 3-dB penalty is associated to the multiplexing of the 4 high-speed data streams at the Tx into a single fiber. Since the joint availability of pulses from different modulators at the same demodulator is of little benefit in the digital communication architecture described here (as opposed to previously explored on-chip ANN architectures), stacking of ODDM super-channels (whole combs) with conventional coarse wavelength division multiplexing (CWDM) [20] might scale more favorably.

The optical power penalties described above also have to be put in relation with the very substantial reduction in modulator count, from 4 to 1 in Fig. 1(b) and 16 to 4 in Fig. 1(c). Given the high power consumption of non-resonantly enhanced modulators and their drivers, this represents a very substantial improvement even when benchmarking against distributed drivers [28], [29]. At the same time, the complexity of the Rx is not substantially increased compared to using a large number of parallelized low-baud-rate links. A more quantitative discussion is included in Section IV after numerically establishing the required optical power levels.

Importantly, a communication system as described here would benefit from generating the comb outside of the package and distributing it over a large number of chips. For one, laser remoting appears a general trend with co-packaged [30] and in-package [2] optics, motivated both by facilitating heat sinking and increasing reliability. Moreover, generating high-quality optical combs, while possible in package [31], [32], generally requires amplification and spectral shaping, the overhead of which might be better shared between a larger number of links. It is noteworthy that the proposed communication scheme thus significantly reduces in-package power consumption (modulators and their drivers) at the expense of out-of-package power consumption (conversion losses during comb generation and higher optical power levels required to overcome link budget penalties). This can be a desirable trade-off, as one of the factors limiting the scaling of current systems is the in-package heat sinking capability.

It should also be mentioned that in addition to the abstract power penalty trade-offs described here, the practical implementation of the system topologies described above are highly dependent on the required comb power levels not exceeding the damage threshold at the optical input port of the chip. In Section IV, we are analyzing the configuration represented by Fig. 1(b) in detail under this aspect.

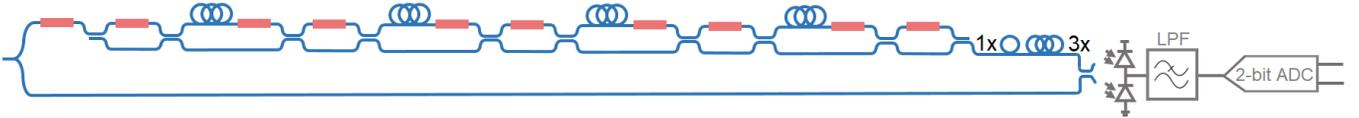

**Fig. 4.** Schematic of a complete demodulator comprising the 5-tap FFE functionality. Fixed delays are combined with a PPS allowing to access different superpositions of delays and configuring the equalizer. Corresponding pulse timing diagrams can be found in Fig. 5.

### III. Equalization Concept

An essential feature of this network concept is its ability to optically equalize signals. This arises from a generalization of the optical sampling occurring in the demodulator. Instead of sampling a single ODDM channel with a single reference pulse, multiple channels can be jointly sampled, weighted and summed by embedding the PPS already introduced in Section II into one of the demodulator arms, as shown in Fig. 4. It delays the incoming pulses by different amounts of time and superimposes them with programmable weights. Each of these programable delays correspond to the reception of a different ODDM channel, which in turn corresponds to a sample of the modulator output taken at a different sampling time. The weights determine the coefficients used in the summation – the very function of a feed forward equalizer (FFE). A fifth unit is added to the PPS to obtain a 5-tap FFE. By delaying the modulated pulses contributing to the interference signal (top branch) rather than the reference branch, the timing of the generated photocurrent pulses is not impacted, so that the following electronics can all be clocked by a single 50 GHz clock, which is essential not to increase the power consumption of the electronic circuitry. After the BPD, a switched integrator (1/50 GHz = 20 ps integration time) or a low-pass anti-aliasing filter (25 GHz cutoff) ensure that the photocurrents are integrated / averaged over the 50 GBaud repetition time of the comb source.

Figure 5 shows exemplary pulse sequences at the end of the two demodulator branches, for different states of the PPS corresponding to incremental delays between 0 to $12\tau_0$, in increments of $3\tau_0$. These are added to the additional delays outside of the PPS, $1\tau_0$ to synchronize with the reference and an additional $3\tau_0$ specific to the demodulator for the ODDM channel of index $n = 1$, that has been exemplarily chosen for illustrative purposes. These ensure that when the PPS introduces a mid-range delay of $6\tau_0$, the demodulator is equivalent to that of Fig. 2, corresponding to the reception of samples from this ODDM channel. The additional delay programmability enables the reception of samples up to two steps before or after the nominal one. Weights are independently optimized for each of the four demodulators, analogous to the approach used in electronic signal processing [33], by training the PSs in the PPS. The resulting configurations create superpositions of the single delay states illustrated in Fig. 5.

As a consequence of the equalization being done in the optical domain, the high-speed, high-resolution ADC typically required in these types of systems [34] can be replaced by simple 4-level thresholding equivalent to a 2-bit ADC. This further contributes to reducing the complexity and power consumption of the Rx electronics, in addition of reducing the CDR rate, removing THAs and buffers, and shifting the starting point of the deserialization to an already much reduced baud rate.

As in electronic systems, equalization comes at a cost in terms of the ratio between the recovered signal strength and the required signal excursions at individual taps. Since here the signal strength recovered at the individual taps is link budget limited (the reference pulse power is spread across the four taps), heavy equalization results in a reduction of the recovered differential photocurrent.

One of the practical challenges in the implementation of such a system results from the possibility of the Tx and Rx PICs being at different temperatures. As described above, the power envelope of the pulse train generated by the pulse interleaver is insensitive to temperature. However, the phase of the generated pulses is not. Since the photocurrent is generated by means of coherent interaction between these and the reference pulse, such phase drift will impact the weights implemented inside the FFEs, unless it is identical in both the Tx and Rx PPSs. Fortunately, as already described in Section II, such a phase drift can be jointly corrected for all the pulses with a single control signal, so that the FFEs do not need to be completely retrained.

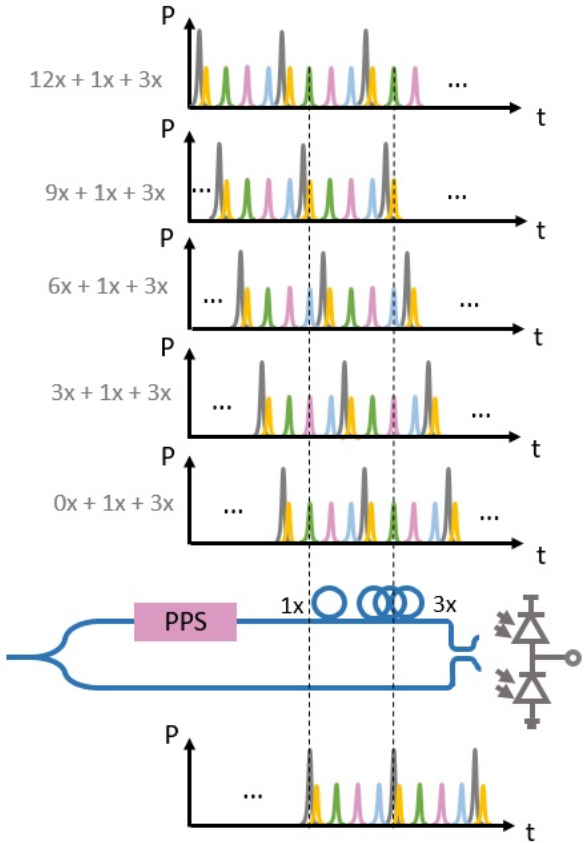

**Fig. 5.** Pulse timing diagrams for the demodulator / FFE aimed at the ODDM channel of index $n = 1$, for PPS states creating pure delays in the range $0, 3\tau_0, \ldots, 12\tau_0$. After training, the PPS creates optimized superpositions of the timing diagrams shown for the upper branch.

## IV. NUMERICAL MODELING

Table 1 summarizes the assumptions made for modeling the network represented in Fig. 1(b). Since the power handling capability is essential at the Tx, waveguides are assumed to be implemented in a high-quality silicon nitride layer (SiN), that can carry much more power than silicon [35], and modulators to be implemented as LNOI on SiN [7]. At the Rx, SiN waveguides can also be utilized for the initial splitter network before light is transferred to silicon-based waveguides [36] and to germanium waveguide photodetectors [37]. The average power in the Rx silicon waveguides and photodetectors is estimated to be below 3 mW and 1 mW, respectively, which is low enough not to induce significant nonlinearities in the waveguides [38] or the photodetectors [39].

The methodology used to model the network and its noise characteristics follows that of [16]. In particular, thermal noise, shot noise, relative intensity noise, as well as the optical and RF linewidths of the comb source are taken into account. Prior to evaluating the network performance, each of the four demodulators are trained with a random sequence of 50 symbols. A weighted error function considers both the accuracy of the rescaled reconstructed signal as well as the magnitude of the rescaling factor. Both are important, since the O-E bandwidth induced distortion must be corrected, but the final signal strength also has to stay above the noise floor.

Figure 6 shows modeling results for five different scenarios in the form of histograms of the recovered samples, aggregated over all four demodulators. The first one ignores Tx O-E bandwidth limitations and assumes the comb power in the fiber leading to the Tx PIC to be 250 mW in a back-to-back link. The second scenario corresponds to the same comb power with the driver and modulator cutoff frequencies taken into account, but without optical FFE, in which case the signal quality can be seen to have catastrophically degraded. In the following ones, FFE is implemented and trained, with comb power levels of 200 mW (scenario 3) and 500 mW (scenario 4). In scenario 5, noise sources are turned off, as a reference.

In all cases, signals applied to the phase modulators are assumed to be predistorted to account for the sine-shaped demodulator transfer function, but without any pre-emphasis addressing the driver + modulator bandwidth limitation.

In order to extract a signal quality from the histograms, we define the Q-scale as the worst case of $(I_3 - I_2)/(\sigma_3 + \sigma_2)$, $(I_2 - I_1)/(\sigma_2 + \sigma_1)$ and $(I_1 - I_0)/(\sigma_1 + \sigma_0)$, with $I_3 - I_2$, $I_2 - I_1$ and $I_1 - I_0$ the upper, middle and lower eye openings determined by the worst-case traces and $\sigma_i$ their respective noise levels. Scenario 1 (250 mW, no E-O bandwidth limitations) results in a Q-scale of 7 sufficient to achieve error free operation. Scenario 2 (250 mW, realistic E-O bandwidth, no FFE) shows a high overlap between the 4 peaks of the histogram, illustrating the need for FFE. Scenario 3 (200 mW, realistic E-O bandwidth with 5-tap optical FFE) features a Q-scale of 3.5, sufficient to obtain a symbol error rate (SER) below $4.8 \times 10^{-4}$ compatible with the 802.3bs IEEE standard for 400-gigabit Ethernet. Scenario 4 (500 mW, realistic E-O bandwidth with 5-tap optical FFE) is again error free. When noise sources are completely turned off, the Q-scale goes to > 16, showing that the optical FFE very adequately compensates for the Tx induced distortion. Figure 7 illustrates the distortion induced by the Tx and the quality of the FFE compensated signal, by comparing initial (Tx In) and recovered (Rx Out) samples for scenarios 2 (no FFE) and 5 (FFE).

**Table 1.** Assumptions used to model the network shown in Fig. 1(b). References used to validate the feasibility are also included in the table for each assumption.

| Specification | Symbol | Value | Reference |
|---|---|---|---|
| Driver voltage swing | $V_{dd}$ | 2.8 V | [10] |
| Driver cutoff frequency | $f_{c,dr}$ | 70 GHz | [10] |
| Modulator $V_\pi$ (for single phase shifter) | $V_\pi$ | 4.6 V | [6] |
| Modulator cutoff frequency | $f_{c,mod}$ | 80 GHz | [6] |
| Modulator insertion loss | $IL_{mod}$ | 0.5 dB | [6] |
| Comb FSR | FSR | 50 GHz | [40] |
| Comb shape | | sech$^2$ | [40] |
| Comb FWHM (single link) | FWHM | 9.6 nm | [40] |
| Average power in fiber, EDFA input | $P_{EDFA,In}$ | 3 mW | [41] |
| Average power in fiber, PIC input | $P_{PIC,In}$ | <500 mW | [35] |
| EDFA noise figure | NF | 5 dB | Typical EDFA |
| Optical linewidth | $\Delta\nu_{opt}$ | 1 MHz | Inherited from pump |
| RF linewidth | $\Delta\nu_{RF}$ | < 1 kHz | [42] |
| Relative intensity noise | $S_I$ | -140 dBc/Hz | Typical DFB pump |
| Input referred TIA thermal noise PSD | $S_{th}$ | 15 $pA/\sqrt{Hz}$ | [37] |
| PIC to fiber losses (per interface) | $IL_{EC}$ | 3 dB | Estimate |
| Excess splitter loss | $IL_{spl,ex}$ | 0.5 dB | Estimate |
| Excess pulse interleaver loss | $IL_{int}$ | 3 dB | Estimate |
| Excess distribution network loss | $IL_{dis}$ | 0.5 dB | Estimate |
| Excess demodulator loss | $IL_{dem}$ | 3 dB | Estimate |
| Waveguide routing losses | $IL_{rtg}$ | 2 dB | Estimate |
| Fiber dispersion parameter | D | 18 ps/nm/km | SMF28 |

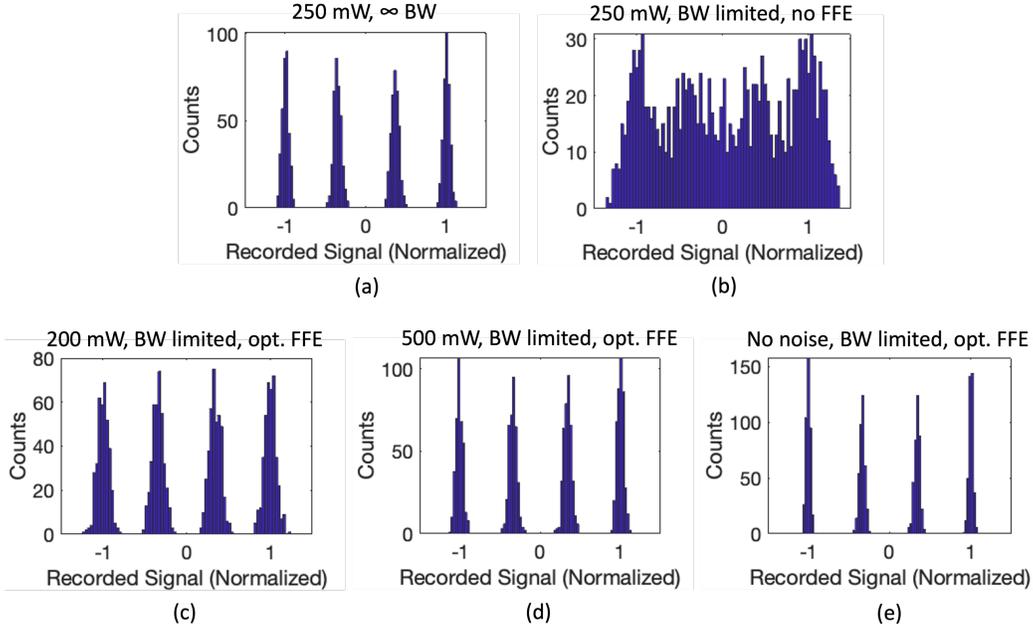

**Fig. 6.** Histograms of the recorded samples aggregated over the four demodulators for 5 different scenarios.

The required 200 mW input power for 400 Gb/s transmission seen in scenario 3 is relatively high given the low wall-plug efficiencies of erbium-doped fiber amplifiers (EDFAs), as required to boost the comb power to the necessary level, which are in the range of 3-10% for this power class [43]. The required power to boost the comb has, however, to be put in relation with the power savings resulting from using a single high-speed TW-phase-modulator instead of driving four independent 50-GBd silicon modulators. In all-silicon technology, the typical Tx power consumption is above 5 pJ/bit, excluding laser power, for non-resonantly enhanced devices [44]-[46]. In comparison, the < 1 W of the high-speed driver considered here [10] results in a comparatively low in-package power consumption below 2.5 pJ/bit, which may be further halved given that a single high-speed phase shifter is needed. Since the thermal phase shifters required in the demodulators are only embedded into passive devices that do not need RF connectivity via metal lines acting as heat pipes, they can be made very efficient by undercutting the silicon [47]. In the Tx, in which the pulse interleaver has to be implemented in SiN, a robust untuned architecture as shown in Fig. 3(a) might be preferred.

We further investigated the effect of GVD by modeling links corresponding to scenario 4 (500 mW, O-E bandwidth limitations, FFE turned on) with different fiber lengths. The Q-scale drops below 3.4, the limit to maintain the target SER, beyond 40 meters. When the FFEs are retrained for every fiber length, the Q-scale stays above 3.4 up to a distance of 66 meters. It could be hypothesized that the ability of the FFEs to mitigate

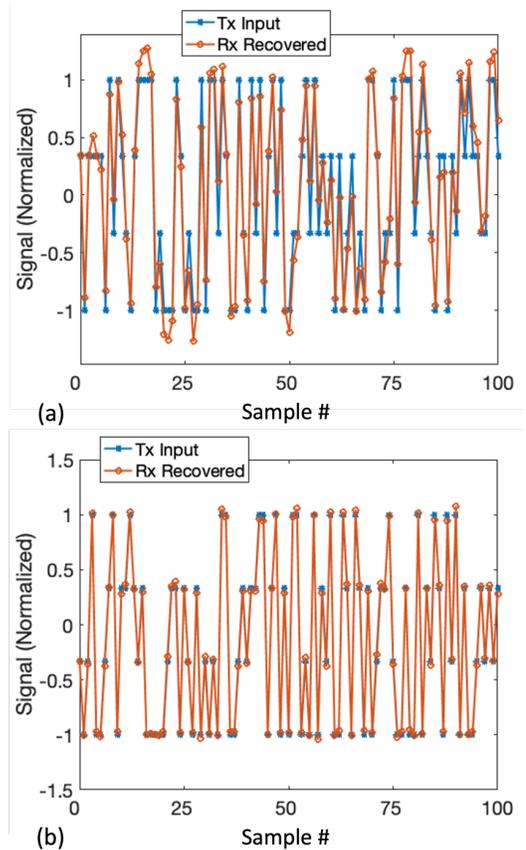

**Fig. 7.** Samples input at the Tx (blue) and recovered at the Rx (red) for scenarios 2 (a) and 5 (b), illustrating the Tx induced distortion and the quality of the FFE compensated signal.

GVD would be even higher in the absence of Tx distortion, as they would then be free to use all available degrees of freedom for the former. This was checked and proved, however, not to be the case. We attribute this to GVD creating crosstalk between all the ODDM channels (see Appendix II), but only a subset of these, that have an index corresponding to a multiple of 3, being available to the evaluated FFE.

Since the full width at half maximum (FWHM) of the required comb is 9.6 nm (corresponding to 2x the minimum comb width required in case of a perfect square spectrum – 12 50-GHz channels spanning 4.8 nm), crosstalk between super-channels would be small if stacked on a 20 nm CWDM grid.

## Conclusion

We have proposed and modeled an orthogonal delay division multiplexed communication system distributing information across the entire spectrum of a frequency comb in an orthogonal manner. A theoretical framework covering both dispersion and aliasing in the optical domain is introduced. The required power budget, in conjunction with the damage threshold of the Tx input port in which the high-power comb is injected, drives the complexity of the systems that can be practically achieved.

We model in detail the concrete example of the transport of 400 Gb/s across a single CWDM channel with a single LNOI modulator at the Tx and a single fiber, taking all noise sources present in the system into account. The data is optically deserialized and equalized at the receiver, greatly reducing the in-package power consumption compared to a high-baud-rate link with all-electronic processing. The achievable fiber length compatible with a $4.8 \times 10^{-4}$ symbol error rate is 66 meters, comparable with the performance of conventional IM/DD modulation schemes at such high baud rates.

## Appendix I: Theoretical Framework

Figure 9(a) illustrates ODDM carriers in the form of a dispersed comb, with comb line $q$ having a random initial phase $\theta_q$ in addition to the deterministic phase $\omega_q n\tau_0$ arising from the delay $n\tau_0$. $n$ is the index of the ODDM carrier and is an integer between 0 and $Q-1$, with $Q$ the number of comb lines, and $\omega_q = \omega_0 + q\delta\omega$ is the angular frequency of the comb line. $\delta\omega$ and $\delta\nu$ represent the FSR of the comb as an angular / ordinary frequency. The time delay $\tau_0$ is chosen to be $1/\Delta\nu$, as explained in the main text, which results in $\omega_q n\tau_0 = \omega_0 n\tau_0 + 2\pi qn/Q$. The incremental phases $2\pi qn/Q$ can be seen to correspond to the coefficients of an inverse discrete Fourier transform (IDFT).

If two such subcarriers are fed through a 2-by-2 DCS followed by a BPD, as occurring in the demodulators, they will not result in a differential photocurrent unless they have the same ODDM carrier index (same $n$), provided the differential photocurrent is integrated over a duration $UI = 1/\delta\nu$ or low-pass filtered in the RF domain with an anti-aliasing filter of cutoff frequency below $\delta\nu/2$ [16]. Both these conditions essentially model the function of an ADC operating at a clock rate of $\delta\nu$, here 50 GHz.

Such carriers can thus be used to carry independent information. In the general case, this holds true even for fully dispersed pulses for which the initial phases $\theta_q$ are completely random, i.e., this is not an OTDM system in the usual sense [16]. This is for example the case in the system depicted in Fig. 1(a), in which signals generated by four independent modulators are mapped to different ODDM channels. As a limitation, when a single ODDM carrier is fed through each modulator, signal frequencies are limited to stay below $\delta\nu/2$ for dual-sided modulation and $\delta\nu$ for single-sided modulation to avoid aliasing, compare Figs. 9(b) and 9(c). In Fig. 9(c), the high frequency content above $\delta\nu/2$ modulated onto the comb line of index $q$ is mapped to the recorded frequency ranges around comb lines $q-1$ and $q+1$, and thus aliased to a smaller RF frequency after demodulation.

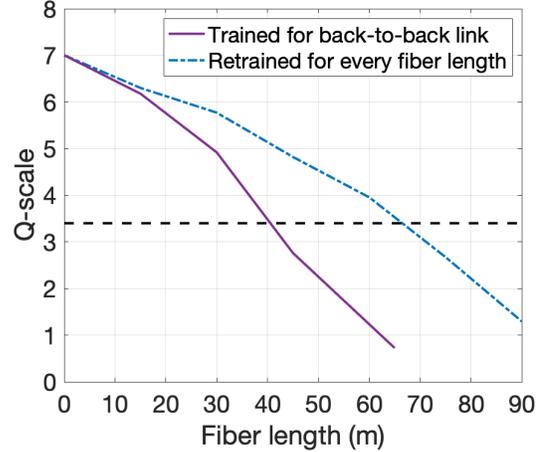

**Fig. 8.** Q-scale for scenario 4 (500 mW with O-E bandwidth limitations and FFE turned on) for different fiber lengths, assuming a fixed set of FFE coefficients (purple curve) and retrained FFEs (blue curve).

This problem can be solved by introducing redundancy in the system and sampling the signal multiple times with different clock phases, as commonly done in high-speed oscilloscopes in the electrical domain [48]. Here, this is achieved by feeding several ODDM carriers through the same modulator, as shown in Fig. 1(b). Sending $N$ ODDM carriers through the modulator, that are delayed relative to each other in increments of $UI/N$, allows increasing the signal bandwidth by a factor $N$ without incurring aliasing after demodulation and post-processing. This is illustrated in Fig. 9(d) for the case $N=4$ implemented in Fig. 1(b).

In the general case, a practical difficulty then arises from the fact that high-frequency RF components ($> \delta\nu/2$) are mapped to other ODDM channels, depending on the set of initial phases $\theta_q$. This precludes architectures such as shown in Fig. 1(b) in which the reference pulse is sent through the same fiber using a dedicated ODDM channel that needs to remain free of crosstalk, or architectures such as shown in Fig. 1(c) in which independent signals generated by different modulators are mapped to different sets of ODDM carriers.

To address this, we utilize undispersed pulses at the input of the modulators, corresponding to the phases $\theta_q$ being all equal to each other. As shown in Appendix III, this results in ODDM channels being mapped to themselves while being aliased, so that the channel crosstalk problem does not occur in back-to-back configuration. In the time domain, sending undispersed pulses through the modulator corresponds to an optimum sampling operation, which is also the information required by a conventional FFE. With the specific example of a square shaped spectrum, as followed in these appendices, this is equivalent to sampling with orthogonal, sinc shaped pulses [21].

In the time domain, the initial data stream from a given modulator can then be recovered by interleaving the signals recorded from the $N$ corresponding demodulators. In the frequency domain, the original spectrum can also be reconstructed from linear superpositions or the recorded spectra, in which case the data processing would reduce to that followed for a filter-less optically enabled ADC reported in [49].

However, while we perform the sampling operation with undispersed pulses, these can disperse thereafter while propagating through the fiber without significantly reducing the demodulated signal quality, provided the GVD stays below an allowable bound. Importantly, pulses corresponding to different ODDM channels are allowed to overlap with each other as a consequence of GVD, as long as they stay within the same UI. This is significant, since many pulses are close packed within one UI. For example, in Figs. 1(b) and 1(c), there are respectively 4 and 16 pulses inside one UI. This is a significant difference from IM/DD, in which interleaved pulses are required to stay disjoint at the Rx [20].

The reason for this is illustrated by Fig. 10. Figure 10(a) shows the phase error induced by GVD across the spectrum of a modulated ODDM channel. The dots represent the phase errors applied to the unmodulated comb lines of the reference pulse. Since the recorded photocurrent results from the self-homodyne interference between the modulated ODDM channel and the reference pulse, the phase errors of the latter are subtracted from that of the modulated ODDM channel during O-E conversion. This results in the diagram shown in Fig. 10(b). The recorded photocurrent results from a summation of each of the spectral slices shown there, with each of these down-converted to the baseband by heterodyne mixing with the nearest reference comb line. The spectral slices shown in Fig. 10(b) are thus all down-converted to DC, but inherit a phase $2\pi q(n - n_R)/Q$, with $n$ the modulated ODDM channel index and $n_R$ the ODDM channel index of the reference. These phases are also responsible for the orthogonality conditions, i.e., a photocurrent is obtained only if $n = n_R$ after time shifting of the signal / reference inside the demodulator. Other than for this phase difference, the $Q$ signals superposed at the BPD are identical in the absence of GVD.

GVD related non-idealities primarily arise from the slope of the frequency dependent phase error now depending on the index $q$ of the corresponding spectral slice. In other words, the spectral slices propagate with different group velocities through the fiber. If we neglect the curvature of the phase error within a given slice, this means that the corresponding signals arrive at different times at the BPD, but are otherwise unchanged. This is a reasonable approximation, as even for Fig. 1(b) that interleaves only four pulses in the time domain, there are at least 12 ODDM channels and thus at least 12 spectral slices, and the GVD across one slice is much smaller than the GVD across the entire spectrum. The different time delays incurred by the signals while propagating through the fiber are irrelevant, so long as they still fall within the same UI in case of an integrating ADC (the logic is the same for an anti-aliasing filter that essentially averages the signal over its underlying time constant). A more rigorous mathematical derivation of the sensitivity to GVD is reported in Appendix II.

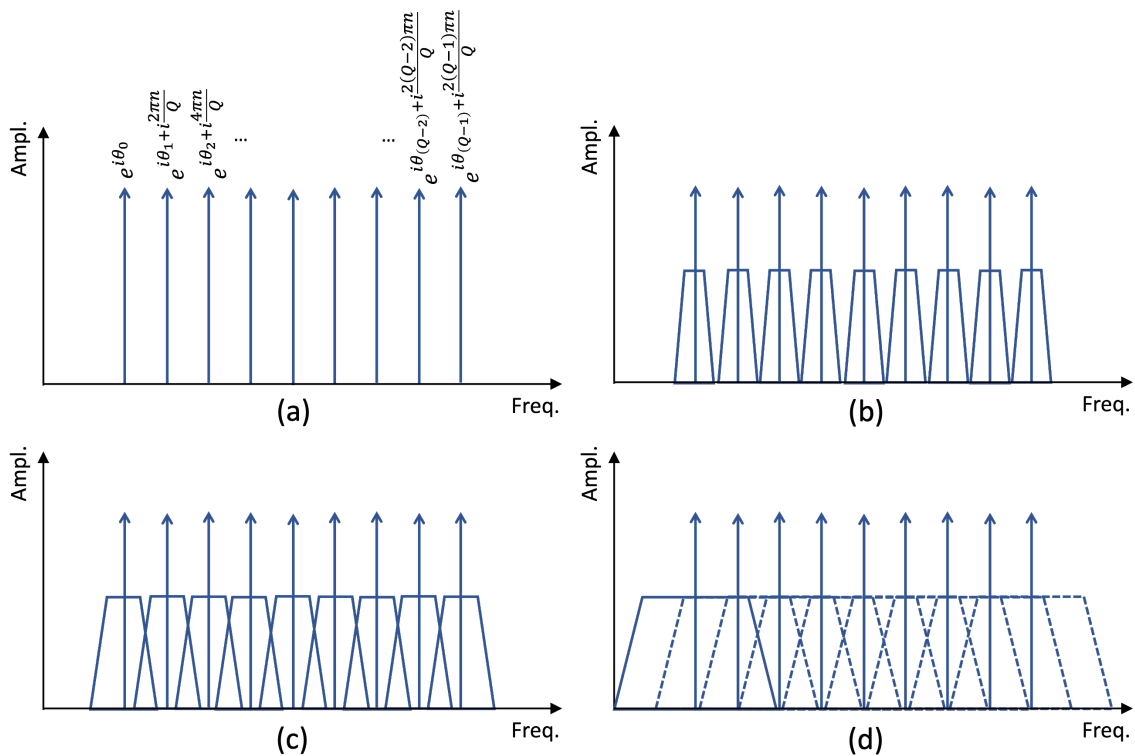

**Fig. 9.** Illustration of (a) ODDM carrier of index $n$ with initial phases $\theta_q$, (b) dual-sided modulation with an RF frequency content below $\delta\nu/2$, (c) aliasing when this frequency content is above $\delta\nu/2$ and (d) the spectra generated in the system shown in Fig. 1(b) with $N = 4$ interleaved pulses and an RF frequency content up to $2\delta\nu$.

We can thus see that the sensitivity to GVD depends on both the initial pulse width, that dictates how fast the pulse broadens, and the UI, that dictates how much broadening is allowable. This is in contrast to an IM/DD system in which the pulse width determines both the rate at which pulses broaden and the amount by which they are allowed to broaden, and is a significant advantage. In terms of GVD tolerance (the link length decreases as the inverse of the number of interleaved channels), this system is between a fully serialized IM/DD link (quadratic decrease with the baud rate) and a WDM link, in which the achievable link length is independent on the number of stacked channels. In the case of Fig. 1(b), the factor $N = 4$ improvement in GVD tolerance is largely offset by the penalty arising from the required number of ODDM channels being 3x more than the number of actually utilized ones to enable transport of the reference pulse through the same fiber. As the system is being scaled up to larger channel counts, the improvement would, however, be more substantial.

APPENDIX II: ORTHOGONALITY CONDITIONS AND DISPERSION

We are assuming here and in the following that the 2-bit ADC in Fig. 2 is implemented with an anti-aliasing filter with a bandwidth corresponding to half the comb FSR. In that case, the differential photocurrent obtained from feeding two ODDM carriers, with indices $n$ and $n_R$ and comb line specific phases $\theta_q + 2\pi nq/Q + \varphi_n$, $\theta_q + 2\pi n_R q/Q$, through the two inputs of a DCS, followed by a BPD and the anti-aliasing filter, is

$$I_d = \frac{2R\sqrt{P_n P_{n_R}}}{Q} Re\left(\sum_{q=0}^{Q-1} e^{i\frac{2\pi q(n-n_R)}{Q}+i\varphi_n}\right) \quad (2)$$

$$= 2R\sqrt{P_n P_{n_R}}\delta_{n,n_R}\cos(\varphi_n)$$

where $\varphi_n$ is some constant phase delay applied to the ODDM carrier of index $n$. $P_n$ and $P_{n_R}$ are the power levels of the ODDM channels fed into the DCS and $\delta_{n,n_R}$ is the Kronecker delta.

We now assume that the ODDM carrier of index $n$ is dual-side modulated with a maximum frequency $\delta\nu/2$, such that comb lines of amplitude $\sqrt{P_n/Q}\, e^{i(\omega_0+q\delta\omega)t+i(\theta_q+2\pi nq/Q)}$ are transformed into

$$A_{n,q} = \sqrt{\frac{P_n}{Q}} \cdot \int_{-\delta\omega/2}^{\delta\omega/2} T(\omega)e^{i\omega t}d\omega \cdot e^{i(\omega_0+q\delta\omega)t+i(\theta_q+2\pi nq/Q)} \quad (3)$$

with $T(\omega)$ the Fourier transform of the applied signal. Equation (2) then becomes

$$I_d = 2R\sqrt{P_n P_{n_R}}Re\left(e^{i\varphi_n}\int_{-\delta\omega/2}^{\delta\omega/2}T(\omega)e^{i\omega t}d\omega\right)\delta_{n,n_R} \quad (4)$$

i.e., the orthogonality is maintained if $n \neq n_R$ and the signal is recovered if $n = n_R$. In the presence of GVD, the situation is more complex and Eq. (2) turns into

$$\frac{2R\sqrt{P_n P_{n_R}}}{Q} Re\left(\sum_{q=0}^{Q-1} e^{i\frac{2\pi q(n-n_R)}{Q}+i\varphi_n}\right.$$
$$\left.\cdot \int_{-\delta\omega/2}^{\delta\omega/2} T(\omega)e^{i\omega t+i\beta_2 q\omega\delta\omega L+i\frac{1}{2}\beta_2\omega^2 L}d\omega\right) \quad (5)$$

where $\beta_2$ describes the GVD of the fiber and $L$ is its length.

In (5), the term $\exp\left(i\frac{1}{2}\beta_2\omega^2 L\right)$ leads to broadening of the received current pulse, but does not compromise the orthogonality conditions since it is common for all the terms. $\exp(i\beta_2 q\omega\delta\omega L)$ on the other hand corresponds to a group delay that depends on $q$ and compromises orthogonality. To analyze this further, we assume that this term is the dominant non-ideality, since $q\delta\omega \gg \omega$ (with $\omega$ bounded by $\delta\omega$), and that the applied RF signal is a pure tone of frequency $\omega_{RF}$. In case of small-signal phase modulation, $T(\omega)$ then reduces to $\delta(0) + ia\delta(\omega_{RF}) - ia\delta(-\omega_{RF})$, with $a$ a real-valued number describing the side-band amplitude depending on the modulation strength. For $n \neq n_R$, (5) then reduces to

$$I_d = \frac{2R\sqrt{P_n P_{n_R}}}{Q} Re\left(\sum_{q=0}^{Q-1} e^{i\frac{2\pi q(n-n_R)}{Q}+i\varphi_n} \cdot \left(1 + iae^{i\omega_{RF}t+i\beta_2 q\omega_{RF}\delta\omega L} - iae^{-i\omega_{RF}t-i\beta_2 q\omega_{RF}\delta\omega L}\right)\right) =$$

$$\frac{2R\sqrt{P_n P_{n_R}}}{Q} Re\left(iae^{i\varphi_n+i\omega_{RF}t}\frac{1-e^{i\beta_2 Q\omega_{RF}\delta\omega L}}{1-e^{i\frac{2\pi(n-n_R)}{Q}+i\beta_2\omega_{RF}\delta\omega L}} - iae^{i\varphi_n-i\omega_{RF}t}\frac{1-e^{-i\beta_2 Q\omega_{RF}\delta\omega L}}{1-e^{i\frac{2\pi(n-n_R)}{Q}-i\beta_2\omega_{RF}\delta\omega L}}\right) \approx$$

$$\frac{2R\sqrt{P_n P_{n_R}}}{Q} Re\left(iae^{i\varphi_n+i\omega_{RF}t}\frac{1-e^{i\beta_2 Q\omega_{RF}\delta\omega L}}{1-e^{i\frac{2\pi(n-n_R)}{Q}}} - iae^{i\varphi_n-i\omega_{RF}t}\frac{1-e^{-i\beta_2 Q\omega_{RF}\delta\omega L}}{1-e^{i\frac{2\pi(n-n_R)}{Q}}}\right) =$$

$$\frac{2R\sqrt{P_n P_{n_R}}}{Q} Re\left(iae^{i\varphi_n-i\frac{2\pi(n-n_R)}{2Q}}\frac{\sin\left(\frac{1}{2}\beta_2 Q\omega_{RF}\delta\omega L\right)}{\sin\left(\frac{2\pi(n-n_R)}{2Q}\right)} 2\cos\left(\omega_{RF}t + \frac{1}{2}\beta_2 Q\omega_{RF}\delta\omega L\right)\right)$$

(6)

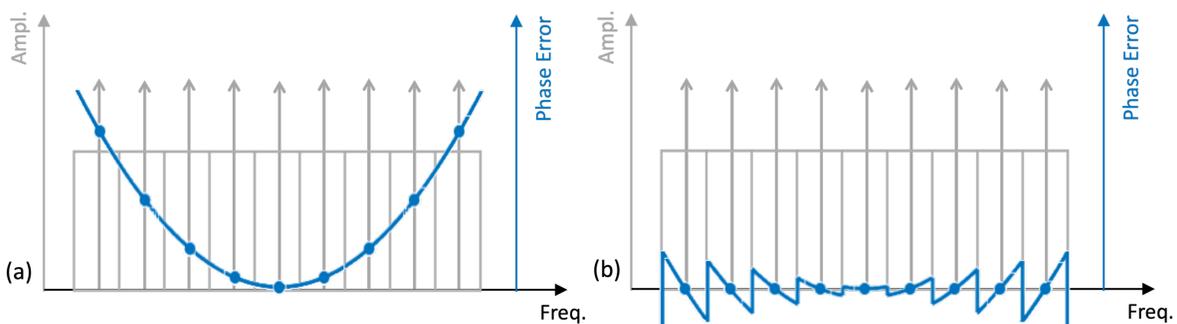

**Fig. 10.** (a) Illustration of the phase error (blue curve) accrued across the modulated ODDM channel spectrum (in grey), as well as the phase error for each of the comb lines of the reference pulse (blue dots). (b) Residual phase error after subtracting the phase error of the nearest reference comb lines from the modulated ODDM channel spectrum. The different slices can be inferred to propagate with different phase velocities through the fiber, as determined by the slope of the residue.

It is thus apparent that the error in the signal rejection grows as $\frac{1}{2}\beta_2 Q\omega_{RF}\delta\omega L = \frac{1}{2}D\lambda_0^2 \omega_{RF}L/\tau_0 c_0$. Since $\omega_{RF} < \pi\delta\nu$, this quantity is also below $\frac{1}{2}D\lambda_0^2\pi L/\tau_0 UIc_0$, which is the result already inferred in a more qualitative way in Appendix I: The sensitivity to GVD depends on the pulse width $\tau_0$ and the initial pulse repetition time defined as one UI.

APPENDIX III: CHANNEL MAPPING WHEN ALIASING IN THE OPTICAL DOMAIN

If the RF modulation frequency is not bounded by half the FSR of the comb, the general expression for a modulated ODDM channel is

$$\sum_{q=0}^{Q-1} A_{n,q} =$$

$$\sqrt{\frac{P_n}{Q}} \cdot \int T(\omega)e^{i\omega t}d\omega \cdot \sum_{q=0}^{Q-1} e^{i(\omega_0+q\delta\omega)t+i(\theta_q+2\pi nq/Q)} =$$

$$\sqrt{\frac{P_n}{Q}} \cdot \sum_{q=0}^{Q-1}\sum_{q'=0}^{Q-1}\int_{-\delta\omega/2}^{\delta\omega/2} T(\omega+(q-q')\delta\omega)e^{i(\omega+(q-q')\delta\omega)t}d\omega \quad (7)$$
$$\cdot e^{i(\omega_0+q'\delta\omega)t+i\left(\theta_{q'}+2\pi nq'/Q\right)} =$$

$$\sqrt{\frac{P_n}{Q}} \cdot \sum_{q=0}^{Q-1}\sum_{q'=0}^{Q-1}\int_{-\delta\omega/2}^{\delta\omega/2} T(\omega+(q-q')\delta\omega)e^{i\omega t}d\omega$$
$$\cdot e^{i\omega_0 t+iq\delta\omega t+i\left(\theta_{q'}+2\pi nq'/Q\right)}$$

This expression can be interpreted as follows: RF frequency slices outside of $[-\delta\omega/2, \delta\omega/2]$ are aliased into it, but at the same time mapped to, in the general case, different ODDM channels, whose comb line specific phases $\varphi(q) = \theta_q + 2\pi nq/Q$ have been replaced by $\varphi(q) = \theta_{q'} + 2\pi nq'/Q$, with $(q'-q)\delta\omega$ the amount by which the RF frequency has been aliased. This change in comb line coefficients, other than for outer comb lines, is essentially a cyclic permutation. A difference from this simple picture arises from the fact that frequency content that is pushed out the spectral range of the comb on the one side is lost, as there are no corresponding reference comb lines to demodulate it, while there is no corresponding frequency content pushed in from the other side of the comb spectrum, for lack of modulated comb lines. In terms of the transformation applied to the ODDM carrier comb line coefficients, this means that coefficients that exit the array on one side during the cyclic permutation do not reenter it from the other, but are instead padded by zeros. However, these boundary effects become negligible when the number of comb lines significantly exceeds the ratio $\omega_{RF,max}/\delta\omega$, with $\omega_{RF,max}$ the maximum RF modulation frequency, which is the case in the numerical examples reported in this paper.

In the special case of a Fourier transform limited pulse, in which case the coefficients $\theta_q$ are zero, the net result of such a cyclic permutation takes a particularly simple form: Since the comb line phases of the ODDM carrier of index $n$ go as $2\pi nq/Q$ and linearly grow with $q$, the cyclic permutation simply maps the array of coefficients onto itself, with a global phase $2\pi n\delta q/Q$ added to all the coefficients. $\delta q = (q'-q)$ is the integer number of comb FSR by which the frequency has been aliased. This means that instead of creating crosstalk, resulting from mapping one ODDM channel to another, the increased bandwidth of the RF channel only creates aliasing within a given ODDM channel. This is advantageous in the types of architectures explored in this paper, in which different ODDM channels are allocated to different tasks, such as carrying signals from a modulator or from the reference, or such as carrying data from different modulators.

Each of the utilized ODDM channels are aliased with a different set of coefficients forming, together, a unitary matrix that can be inverted, so that the initial frequency content can be recovered [49]. Not surprisingly, these linear superpositions of the aliased higher-frequency RF slices correspond to sampling in the time-domain with interleaved, ODDM-channel-index dependent sampling times. Thus, a simpler way of reconstructing the signal consists in interleaving the recorded samples in the time domain. Leaving them in separate data streams thus corresponds to deserialization, which is the approach followed in this paper.